\documentclass[aps,prb,twocolumn]{revtex4-1}

\usepackage[pdftex]{graphics}
\usepackage{amssymb,amsmath}
\usepackage{nicefrac}
\usepackage{physics}
\usepackage{color}
\usepackage{braket}
\DeclareSymbolFont{boldoperators}{OT1}{cmr}{bx}{n}
\SetSymbolFont{boldoperators}{bold}{OT1}{cmr}{bx}{n}
\edef\bar{\unexpanded{\protect\mathaccentV{bar}}\number\symboldoperators16}

\begin{document}

\def\bra#1{\left<{#1}\right|}
\def\ket#1{\left|{#1}\right>}
\def\expval#1#2{\bra{#2} {#1} \ket{#2}}
\def\mapright#1{\smash{\mathop{\longrightarrow}\limits^{_{_{\phantom{X}}}{#1}_{_{\phantom{X}}}}}}

\title{Analytic continuation of Wolynes theory into the Marcus inverted regime}
\author{Joseph E. Lawrence}
\affiliation{Department of Chemistry, University of Oxford, Physical and Theoretical Chemistry Laboratory, South Parks Road, Oxford, OX1 3QZ, UK}
\author{David E. Manolopoulos}
\affiliation{Department of Chemistry, University of Oxford, Physical and Theoretical Chemistry Laboratory, South Parks Road, Oxford, OX1 3QZ, UK}

\begin{abstract}
The Wolynes theory of electronically nonadiabatic reaction rates [P. G. Wolynes, J. Chem. Phys. {\bf 87}, 6559 (1987)] is based on a saddle point approximation to the time integral of a reactive flux autocorrelation function in the nonadiabatic (golden rule) limit. The dominant saddle point is on the imaginary time axis at $t_{\rm sp}=i\lambda_{\rm sp}\hbar$, and provided $\lambda_{\rm sp}$ lies in the range $-\beta/2\le\lambda_{\rm sp}\le\beta/2$, it is straightforward to evaluate the rate constant using information obtained from an imaginary time path integral calculation. However, if $\lambda_{\rm sp}$ lies outside this range, as it does in the Marcus inverted regime, the path integral diverges. This has led to claims in the literature that Wolynes theory cannot describe the correct behaviour in the inverted regime. Here we show how the imaginary time correlation function obtained from a path integral calculation can be analytically continued to $\lambda_{\rm sp}<-\beta/2$, and the continuation used to evaluate the rate in the inverted regime. Comparisons with exact golden rule results for a spin-boson model and a more demanding (asymmetric and anharmonic) model of electronic predissociation show that the theory it is just as accurate in the inverted regime as it is in the normal regime.
\end{abstract}

\maketitle

\section{Introduction}

The Marcus theory of electron transfer rates\cite{Marcus85} is widely regarded as one of the seminal achievements of theoretical chemistry. However, Marcus theory is not always accurate, because it neglects quantum mechanical (zero point energy and tunnelling) effects in the nuclear motion. It is well established, for example, that tunnelling through the reaction barrier can enhance the electron transfer rate by several orders of magnitude over the Marcus theory prediction in the Marcus inverted regime.

Thirty years ago, Wolynes proposed a simple generalisation of Marcus theory that includes zero point energy and tunnelling effects.\cite{Wolynes87}  Starting from the nonadiabatic (golden rule) limit of the exact quantum mechanical expression for the electron transfer rate coefficient in terms of a time integral of a reactive flux autocorrelation function, he proposed a saddle point approximation to the time integral and argued that the ingredients in this approximation could be extracted from an imaginary time path integral calculation. The resulting theory was clearly applicable to realistic simulations with anharmonic force fields, and it was applied soon afterwards to fully atomistic simulations of both inorganic\cite{Wolynes89,Bader90} and biochemical\cite{Wolynes91} electron transfer reactions in aqueous solution. Cao {\em et al.} have since developed the numerical implementation of the theory in more detail, and suggested a way in which it can be extended beyond the nonadiabatic limit and used to study electron transfer reactions with larger (non-perturbative) electronic coupling strengths.\cite{Cao95,Cao97}

But since then, Wolynes theory seems to have been largely ignored. Most of the modern work in this area seems to be aimed at extending heuristic methods such as centroid molecular dynamics\cite{Cao94a,Cao94b} and ring polymer molecular dynamics\cite{Craig04,Habershon13} to electronically nonadiabatic processes.\cite{Shushkov12,Richardson13,Ananth13,Duke16,Menzeleev14,Kretchmer16} We feel that this is a pity, because Wolynes theory provides a more promising way to calculate electronically nonadiabatic reaction rates than any heuristic approach could ever do. In particular, it is clear how to improve on its approximations. One can in principle go beyond the golden rule limit by including contributions to the correlation function of higher order in the electronic coupling strength,\cite{Wolynes87,Mukamel88} and when the saddle point approximation breaks down one can always use a higher-order phase integral approximation.\cite{Wolynes87,Footnote1}

One possible reason why Wolynes theory has not received the attention it deserves is that there has been some confusion in the literature about its domain of applicability. The imaginary time path integral expression for the reactive flux autocorrelation function at time $t=i\lambda\hbar$ only converges when $\lambda$ is within a strip of width $\beta=1/k_{\rm B}T$. (Most people consider the standard version of the autocorrelation function, for which this strip is $0\le\lambda\le\beta$. Here we shall consider the symmetrically-thermalised version, for which it is $-\beta/2\le\lambda\le\beta/2$.) The key issue is how to deal with situations in which the saddle point lies outside the strip, in a region where the path integral is undefined. For the spin-boson model -- the most basic model of condensed phase electron transfer, for which Marcus theory is exact in the high temperature limit -- the issue arises when the electron transfer is in the Marcus inverted regime.

To illustrate why this is an issue we shall simply give two quotes from a recent paper by Richardson and Thoss,\cite{Richardson14} who have made a number of significant contributions to the theory of electronically nonadiabatic reaction rates (in particular on the nonadiabatic generalisation of the semiclassical instanton approximation,\cite{Richardson15a,Richardson15b} which clearly {\em cannot} be used in the Marcus inverted regime in the form they presented\cite{Footnote2}):\\

 \lq\lq There would however be a problem with [the Wolynes] approach in the Marcus
inverted regime, where the stationary point falls outside the interval $[0,\beta\hbar]$ and
thus has no meaning for path integrals of period $\beta\hbar$."\\

\noindent
and\\

\lq\lq \dots the path integral golden-rule rate formula proposed by Wolynes,
Cao and Voth cannot describe the correct behaviour in the inverted regime $\epsilon>\Lambda$."\\

The purpose of the present paper is to show that there is nothing wrong with Wolynes theory in the Marcus inverted regime that a little (and it turns out to be a very little) bit of analytic continuation cannot cure. Sec.~II summarises Wolynes theory,\cite{Wolynes87} Sec.~III discusses its numerical implementation, and Sec.~IV presents example applications to two model problems for which the exact golden rule rates can be computed for comparison: a spin-boson model with Debye spectral density and a one-dimensional model of electronic predissociation. For both of these models, the analytic continuation of Wolynes theory into the Marcus inverted regime is found to be just as accurate as the theory is in the normal regime.

\section{Wolynes Theory}

The Hamiltonian for two coupled diabatic electronic states is
\begin{equation}
  \hat{H} = \hat{H}_0\dyad{0}{0}+\hat{H}_1\dyad{1}{1}+\Delta\big(\dyad{0}{1}+\dyad{1}{0}\big),
\end{equation}
where $\Delta$ is the electronic coupling (a constant within the Condon approximation\cite{Nitzan06}) and $\hat{H}_i$ is the diabatic Hamiltonian for state $i$:
\begin{equation}
  \hat{H}_i = \sum_{k=1}^f \frac{\hat{p}_k^2}{2m_k} + V_i(\hat{\bf{q}}).
\end{equation}

The rate constant for transfer from state $\ket{0}$ to $\ket{1}$ can be written in terms of the time integral of a thermally symmetrised flux-flux correlation function,\cite{Yamamoto60,Miller74,Miller83}
\begin{equation}
  k(T)Q_{\rm r}(T)=\frac{1}{2}\int_{-\infty}^{\infty} \tr[e^{-\beta\hat{H}/2}\hat{F}(0)e^{-\beta\hat{H}/2}\hat{F}(t)] \mathrm{d}t,
\end{equation}
where 
\begin{equation}
Q_{\rm r}(T)=\tr[e^{-\beta \hat{H}}\dyad{0}{0}]
\end{equation}
is the reactant partition function and 
\begin{equation}
\hat{F}(t)=e^{+i\hat{H}t/\hbar}\hat{F}e^{-i\hat{H}t/\hbar}
\end{equation}
with
\begin{equation}
  \begin{aligned}
    \hat{F} &= \frac{i}{\hbar}[\hat{H},\dyad{1}{1}]= \frac{i\Delta}{\hbar}(\dyad{0}{1}-\dyad{1}{0})
  \end{aligned}
\end{equation}
is a time-evolved reactive flux operator.

In the nonadiabatic limit where $\Delta\to0$, one can use the fact that $\bra{i}e^{-(\beta/2\pm it/\hbar)\hat{H}}\ket{j}=\mathcal{O}(\Delta)$ for $i\neq j$ and $\bra{i}e^{-(\beta/2\pm it/\hbar)\hat{H}}\ket{i}=e^{-(\beta/2\pm it/\hbar)\hat{H}_i}+\mathcal{O}(\Delta)$ to simplify Eq.~(3) and obtain the golden rule expression for the rate constant\cite{Wolynes87}
\begin{equation}
  k(T)Q_{\rm r}(T)= \frac{\Delta^2}{\hbar^2} \int_{-\infty}^\infty c(t)\, \mathrm{d}t,
\end{equation}
where
\begin{equation}
c(t) = \tr_{\rm n}\Big[e^{-(\beta/2+it/\hbar)\hat{H}_0}e^{-(\beta/2-it/\hbar)\hat{H}_1}\Big]
\label{c_t_def}
\end{equation}
and ${\rm tr}_{\rm n}$ denotes a trace over the nuclear degrees of freedom.
Note in passing that $c(t)$ as defined in Eq.~(8) is a convergent representation of an analytic function for $-\beta\hbar/2\leq \Im t \leq \beta\hbar/2$. There is no direct way of evaluating $c(t)$ beyond this strip because $\bra{{\bf q}}e^{+\epsilon\hat{H}_i}\ket{{\bf q}'}$ diverges for Hamiltonians $\hat{H}_i$ that are not bounded from above when $\Re \epsilon>0$. However, once $c(t)$ has been calculated within the strip it can in principle be analytically continued and evaluated along with its derivatives at any nonsingular point in the complex $t$ plane.

Starting from Eq.~(7), Wolynes made three key observations:\cite{Wolynes87} 
\begin{enumerate}
\item The correlation function $c(t)$ is typically either approximately Gaussian (which it is for symmetric electron transfer\cite{Menzeleev11}) or highly oscillatory (which it is in the Marcus inverted regime\cite{Menzeleev11}). In either case, the integral in Eq.~(7) can be evaluated rather accurately using the saddle point approximation.
\item The dominant saddle point is expected to be on the imaginary time axis. (This is certainly the case for the spin-boson model, for which this saddle point moves from $it=0$ for symmetric electron transfer to $it=\beta\hbar/2$ for activationless electron transfer to $it>\beta\hbar/2$ in the Marcus inverted regime.) This suggests letting $t=i\lambda\hbar$ and $c(i\lambda\hbar)=e^{-\beta F(\lambda)}$, deforming the contour of integration in Eq.~(7) to pass through the saddle point at $t_{\rm sp}=i\lambda_{\rm sp}\hbar$ where $F'(\lambda)=0$, and evaluating the time integral using the saddle point approximation
\begin{equation}
    \int_{-\infty}^\infty  c(t) \,  \mathrm{d}t \simeq \sqrt{\frac{2\pi\hbar^2}{-\beta F''(\lambda_{\rm sp})}}e^{-\beta F(\lambda_{\rm sp})}.
    \label{Saddle_Point_Approx}
\end{equation}
\item
The ingredients $F(\lambda_{\rm sp})$ and $F''(\lambda_{\rm sp})$ in this approximation are straightforward to calculate using imaginary time path integral methods when $-\beta/2\leq\lambda_{\rm sp}\leq\beta/2$.
\end{enumerate}

Bringing all this together, one can write the Wolynes (quantum instanton-like\cite{Miller03}) expression for the rate constant as
\begin{equation}
  k(T)Q_{\rm r}(T)\simeq \frac{\Delta^2}{\hbar} \sqrt{\frac{2\pi}{-\beta F''(\lambda_{\rm sp})}}e^{-\beta F(\lambda_{\rm sp})},
  \label{Wolynes_Theory}
\end{equation}
where
\begin{equation}
  F(\lambda) = -\frac{1}{\beta}\ln\Big\{\tr_{\rm n}\Big[e^{-(\beta/2-\lambda) \hat{H}_0}e^{-(\beta/2+\lambda) \hat{H}_1}\Big]\Big\}.
  \label{F_lambda_def}
\end{equation}
Clearly,
\begin{equation}
  F(-\beta/2) = -\frac{1}{\beta}\ln\Big\{\tr_{\rm n}\Big[e^{-\beta \hat{H}_0}\Big]\Big\}\equiv -\frac{1}{\beta} \ln Q_{\rm r}(T),
\end{equation}
and
\begin{equation}
  F(\beta/2) = -\frac{1}{\beta}\ln\Big\{\tr_{\rm n}\Big[e^{-\beta \hat{H}_1}\Big]\Big\}\equiv -\frac{1}{\beta} \ln Q_{\rm p}(T),
\end{equation}
where $Q_{\rm r}(T)$ and $Q_{\rm p}(T)$ are the reactant and product partition functions in the nonadiabatic limit. $F(\lambda)$ is therefore a free energy, which goes from the free energy of the reactants at $\lambda=-\beta/2$ via the free energy of the transition state at $\lambda=\lambda_{\rm sp}$ [see Eq.~(10)] to the free energy of the products at $\lambda=\beta/2$.

\section{Numerical Implementation}

In the region $-\beta/2\leq \lambda\leq \beta/2$, a standard imaginary time path integral discretisation\cite{Tuckerman10} of
\begin{equation}
  c(i\lambda\hbar) = \tr_{\rm n}\Big[e^{-(\beta/2-\lambda)\hat{H}_0}e^{-(\beta/2+\lambda)\hat{H}_1}\Big]
\end{equation}
gives
\begin{equation}
  c(i\lambda_j\hbar) = \frac{1}{(2\pi\hbar)^{nf}} \int \mathrm{d}^{nf}\mathbf{p}\int \mathrm{d}^{nf}\mathbf{q}\, e^{-\beta_n H_j(\mathbf{p},\mathbf{q})},
  \label{c_lambda_PI}
\end{equation}
where $n$ is the number of path integral beads, $f$ is the number of nuclear degrees of freedom, $\beta_n=\beta/n$, and $\lambda_j=j\beta/n-\beta/2$, with
\begin{equation}
  H_{0}(\mathbf{p},\mathbf{q}) = h(\mathbf{p},\mathbf{q}) + \sum_{i=0}^{n-1} V_{0}(\mathbf{q}_i),
\end{equation}
\begin{equation}
  \begin{aligned}
  H_j(\mathbf{p},\mathbf{q}) =&\,\, h(\mathbf{p},\mathbf{q}) + \frac{V_0(\mathbf{q}_0) + V_1(\mathbf{q}_0)}{2} + \sum_{i=1}^{j-1} V_1(\mathbf{q}_i)\\
  & + \frac{V_0(\mathbf{q}_{j}) + V_1(\mathbf{q}_{j})}{2} + \sum_{i=j+1}^{n-1} V_0(\mathbf{q}_i),
\end{aligned}
\end{equation}
for $j=1,\dots,n-1$, and
\begin{equation}
  H_{n}(\mathbf{p},\mathbf{q}) = h(\mathbf{p},\mathbf{q}) + \sum_{i=0}^{n-1} V_{1}(\mathbf{q}_i).
\end{equation}
Here $h(\mathbf{p},\mathbf{q})$ is the usual free ring-polymer Hamiltonian
\begin{equation}
h(\mathbf{p},\mathbf{q}) = \sum_{i=0}^{n-1}\sum_{k=1}^f \left[{p_{i,k}^2\over 2m_k}+{1\over 2}m_k\omega_n^2\left(q_{i,k}-q_{i+1,k}\right)^2\right],
\end{equation}
with $\omega_n=1/\beta_n\hbar$ and $q_{n,k}\equiv q_{0,k}$.\cite{Craig04}

Using the definition of $F(\lambda)$ in Eq.~(11) we have that
\begin{equation}
  F'(\lambda) = -\frac{1}{\beta}\frac{d}{d\lambda}\ln\big[c(i\lambda\hbar)\big],
\end{equation}
and
\begin{equation}
  F''(\lambda) = -\frac{1}{\beta}\frac{d^2}{d\lambda^2}\ln\big[c(i\lambda\hbar)\big],
\end{equation}
from which it is straightforward to show that
\begin{equation}
  F'(\lambda_j) = -\frac{1}{\beta}\Big\langle\big(V_0(\mathbf{q}_0)-V_1(\mathbf{q}_0)\big)\Big\rangle_j,
\end{equation}
and
\begin{equation}
  \begin{aligned}
  F''(\lambda_j) =& -\frac{1}{\beta}\Big\langle\big(V_0(\mathbf{q}_0)-V_1(\mathbf{q}_0)\big)\big(V_0(\mathbf{q}_j)-V_1(\mathbf{q}_j)\big)\Big\rangle_j\\
  &+\frac{1}{\beta}\Big\langle\big(V_0(\mathbf{q}_0)-V_1(\mathbf{q}_0)\big)\Big\rangle_j^2,
\end{aligned}
\end{equation}
where
\begin{equation}
  \langle A \rangle_j =  \frac{\int \mathrm{d}^{nf}\mathbf{p}\int \mathrm{d}^{nf}\mathbf{q}\, A(\mathbf{q})\, e^{-\beta_n H_j(\mathbf{p},\mathbf{q})}}{\int \mathrm{d}^{nf}\mathbf{p}\int \mathrm{d}^{nf}\mathbf{q}\, e^{-\beta_n H_j(\mathbf{p},\mathbf{q})}}
\end{equation}
is the canonical average of the estimator $A({\bf q})$ in the ensemble with partition function $c(i\lambda_j\hbar)$ and Hamiltonian $H_j(\mathbf{p},\mathbf{q})$. When $j=0$, this Hamiltonian is simply the standard ring polymer Hamiltonian on the reactant diabat, and when $j=n$ it is the ring polymer Hamiltonian on the product diabat. But for intermediate values of $j$, beads $j+1$ to $n-1$ are on the reactant diabat, beads $1$ to $j-1$ are on the product diabat, and beads $0$ and $j$ are on the average of the two diabats, as illustrated for a specific case in Fig.~1.

\begin{figure}[t]
\centering
\resizebox{0.65\columnwidth}{!} {\includegraphics{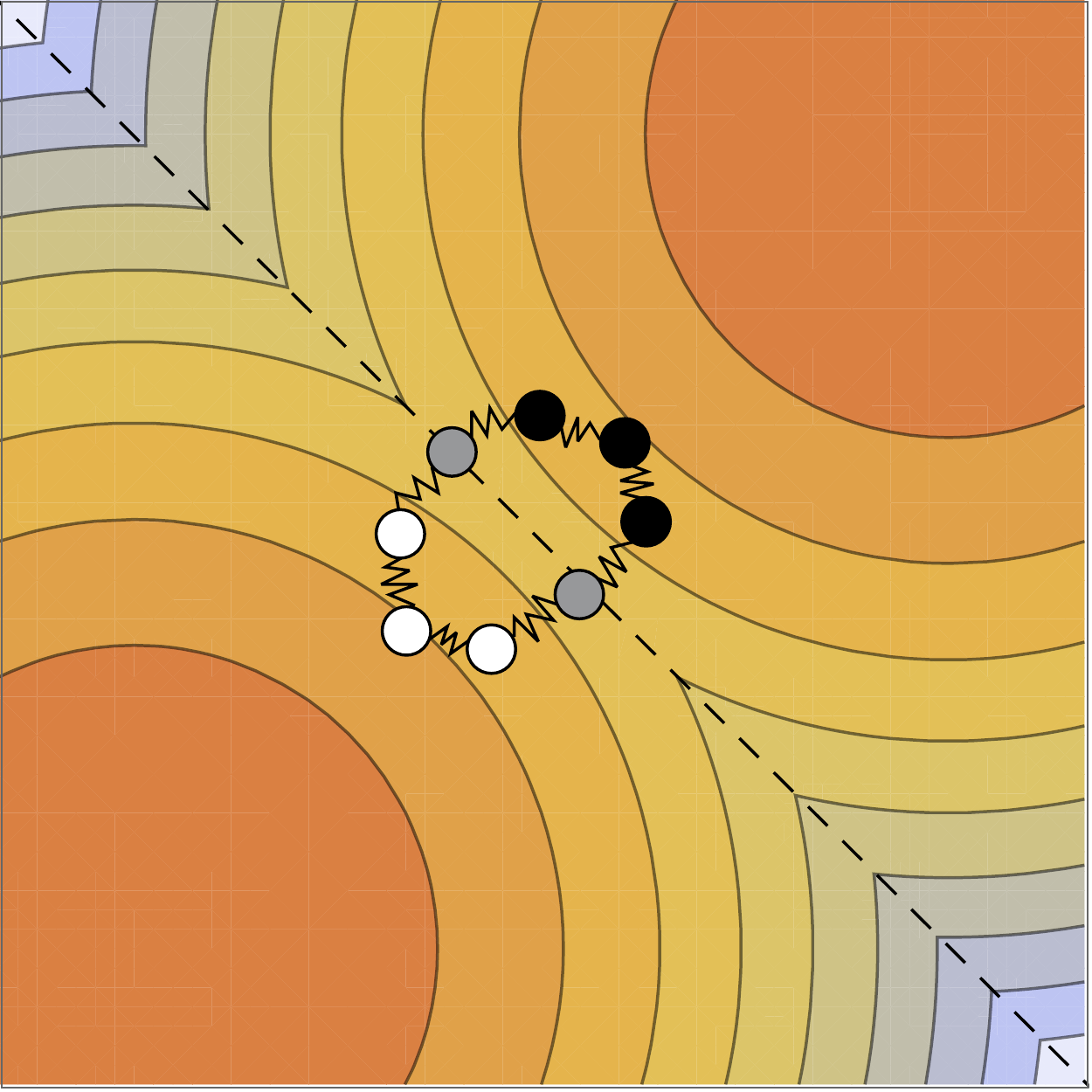}}
\caption{Schematic illustration of the ring polymer Hamiltonian $H_j({\bf p},{\bf q})$ in Eq.~(17) for a problem with $f=2$, $n=8$, and $j=4$. Beads 5 to 7 are on the reactant diabat (white), beads 1 to 3 are on the product diabat (black), and beads 0 and 4 are on the average of the two diabats (grey).}
\label{SB_fits_figure}
\end{figure}

The evaluation of Eqs.~(22) and~(23) requires a separate path integral calculation for each value of $j$. But note that since
\begin{equation}
  \begin{aligned}
  F(\epsilon,\lambda) =& -\frac{1}{\beta}\ln\Big\{\tr_{\rm n}\Big[e^{-(\beta/2-\lambda) \hat{H}_0}e^{-(\beta/2+\lambda) (\hat{H}_1-\epsilon)}\Big]\Big\}\\
  = &\,\, F(\lambda)-\bigg(\frac{1}{2}+\frac{\lambda}{\beta}\bigg)\epsilon,
\end{aligned}
\end{equation}
these path integral calculations only have to be done once to obtain results for all thermodynamic driving forces $\epsilon$. And since the saddle point condition for a given $\epsilon$ is
\begin{equation}
{\partial F(\epsilon,\lambda)\over\partial \lambda} = F'(\lambda)-\frac{\epsilon}{\beta}=0,
\end{equation}
the corresponding value of $\lambda_{\rm sp}$ can be found by solving $\beta F'(\lambda_{\rm sp})=\epsilon$. 

With these observations in mind, the Wolynes rate can be evaluated for any required $\epsilon$ as follows. Combining Eqs.~(10), (12), and~(25) gives
 \begin{equation}
 k(T) \simeq {\Delta^2\over\hbar}\sqrt{2\pi\over -\beta F_{\epsilon}''(\lambda_{\rm sp})}e^{-\beta F_{\epsilon}(\lambda_{\rm sp})},
 \end{equation}
 where we have eliminated the reactant partition function by defining $F_{\epsilon}(\lambda)$ to be the free energy at $\lambda$ minus the free energy of the reactants for a given $\epsilon$,
 \begin{equation}
F_{\epsilon}(\lambda) = F(\lambda)-F(-\beta/2)-\left({1\over 2}+{\lambda\over\beta}\right)\epsilon.
 \end{equation}
It is clear from this that the required quantities are $\lambda_{\rm sp}$, $F_{\epsilon}(\lambda_{\rm sp})$, and $F_{\epsilon}''(\lambda_{\rm sp})$. We begin by assuming a suitable (physically motivated) functional form $\tilde{F}_{\epsilon}(\lambda)$ for $F_{\epsilon}(\lambda)$, and least-squares fitting the parameters in this functional form to the path integral data $\left\{F'(\lambda_j)\right\}$ and $\left\{F''(\lambda_j)\right\}$ at $\epsilon=0$. (The choice of a suitable functional form will be discussed in detail for each of the model problems we shall consider below.) For each required $\epsilon$, we then solve $\tilde{F}'_{\epsilon}(\lambda_{\rm sp})=0$ for $\lambda_{\rm sp}$, and use $\tilde{F}_{\epsilon}(\lambda_{\rm sp})$ and $\tilde{F}_{\epsilon}''(\lambda_{\rm sp})$ in Eq.~(27) to calculate $k(T)$. All of this is straightforward to do if $\tilde{F}_{\epsilon}(\lambda)$ and its derivatives are easy to evaluate. 
 
It is clear that this procedure will work when $\lambda_{\rm sp}$ lies in the range $-\beta/2\le \lambda_{\rm sp}\le \beta/2$, for two reasons. First, one can show by re-writing the right hand side of Eq.~(14) in terms of the eigenvalues and eigenstates of $\hat{H}_0$ and $\hat{H}_1$ and substituting the result into Eq.~(21) that $F''(\lambda)\le 0$ for all $-\beta/2\le \lambda\le\beta/2$. This implies that $F'(\lambda)$ is a monotonically decreasing function of $\lambda$ in this range and that the equation $\beta F'(\lambda_{\rm sp})=\epsilon$ has a unique solution for $\lambda_{\rm sp}$. Secondly, the use of $ \tilde{F}_{\epsilon}(\lambda_{\rm sp})$ and $\tilde{F}_{\epsilon}''(\lambda_{\rm sp})$ in Eq.~(27) is tantamount to interpolating the path integral data to the saddle point when $-\beta/2\le\lambda_{\rm sp}\le\beta/2$, and interpolation is generally quite reliable.

It is less clear that the same procedure will work when $\lambda_{\rm sp}<-\beta/2$, as occurs in the Marcus inverted regime. One reason to suspect that it might work is that the exact $F(\lambda)$ for the spin-boson model [see Eq.~(34) below] satisfies $F''(\lambda)<0$ for all $\lambda$, and not just in the strip $-\beta/2\le\lambda\le\beta/2$. The solution of $\beta F'(\lambda_{\rm sp})=\epsilon$ is therefore unique for this model in both the normal and inverted regimes. Another reason is that the physically-motivated functional form for $ \tilde{F}_{\epsilon}(\lambda)$ for the second (asymmetric and anharmonic) model problem we shall consider below also satisfies $\tilde{F}_{\epsilon}''(\lambda)<0$ for all $\lambda<\beta/2$, and so again gives a unique saddle point in both the normal and inverted regimes. But one cause for concern is that the use of $ \tilde{F}_{\epsilon}(\lambda_{\rm sp})$ and $\tilde{F}_{\epsilon}''(\lambda_{\rm sp})$ in Eq.~(27) is tantamount to {\em extrapolating} the path integral data when $\lambda_{\rm sp}<-\beta/2$, and extrapolation is typically less reliable than interpolation.

\begin{figure}[t]
\centering
\resizebox{0.95\columnwidth}{!} {\includegraphics{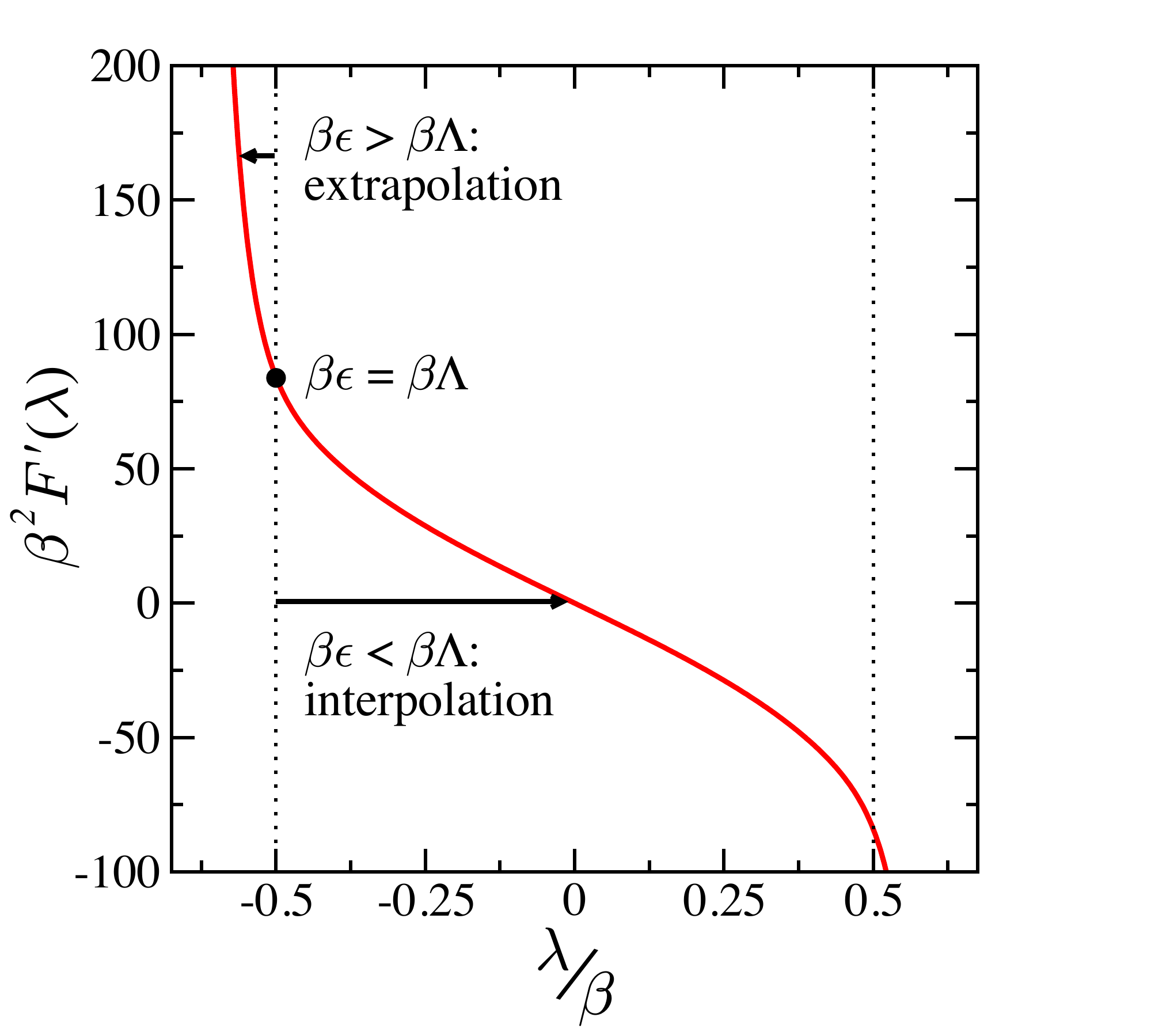}}
\caption{Solutions of $\beta F'(\lambda_{\rm sp})=\epsilon$, for a spin-boson model. Systems in the normal regime, $\epsilon<\Lambda$, are solved by interpolation, while systems in the Marcus inverted regime $\epsilon>\Lambda$ require the fitted curve to be extrapolated. But it does not have to be extrapolated very far beyond the dotted lines within which the path integral data is available, even when $\epsilon=2\Lambda$.}
\label{first_derivative_ext_int}
\end{figure}

This difference between the normal and inverted regimes is illustrated in Fig.~2, with data from the spin-boson model we shall consider in Sec.~IV. Here the reorganisation energy $\Lambda$ is defined as 
\begin{equation}
 \Lambda=\Big\langle\big(V_1(\mathbf{q}_0)-V_0(\mathbf{q}_0)\big)\Big\rangle_0= \beta F'(-\beta/2),
\label{reorgEnergy}
\end{equation}
from which it is clear that the transition to the inverted regime at $\epsilon=\Lambda$ occurs when $\lambda_{\rm sp}=-\beta/2$. (We have already mentioned in Sec.~I that this is the case for the spin-boson model, but Eq.~(29) shows that the result is true more generally.) In the normal regime where $\epsilon<\Lambda$, the solution of $\beta F'(\lambda_{\rm sp})=\epsilon$ is in the range between the dotted vertical lines in Fig.~2, where the path integral data $\{F'(\lambda_j)\}$ and $\{F''(\lambda_j)\}$ has been calculated and fitted to obtain the red curve. In the inverted regime where $\epsilon > \Lambda$, the curve has to be extrapolated to find the solution of $\beta F'(\lambda_{\rm sp})=\epsilon$. But fortunately, for this problem, which has physically reasonable parameters as a model for condensed phase electron transfer, the extrapolation does not extend very far beyond the region where the path integral data is available, even deep in the inverted regime at $\epsilon=2\Lambda$. This suggests that, while extrapolation can in general be problematic, it is on safe ground in this case and can indeed be used in conjunction with Eq.~(27) to calculate Wolynes rate constants in the Marcus inverted regime. 

\section{Results and Discussion}

\subsection{The spin-boson model}

The first system we shall consider is the spin-boson model. Although very simple, consisting of two sets of displaced harmonic oscillators with a constant coupling ($\Delta$), this model contains much of the important physics of condensed phase electron transfer and is commonly used in the study of real physical problems.\cite{Ando01,Blumberger06,Blumberger08} The two diabatic potentials are
\begin{equation}
  V_0(\mathbf{q}) = \sum_{k=1}^{f} \frac{1}{2}m\omega_k^2(q_k+\xi_k)^2,
\end{equation}
and (including the thermodynamic bias)
\begin{equation}
  V_1(\mathbf{q}) = \sum_{k=1}^{f} \frac{1}{2}m\omega_k^2(q_k-\xi_k)^2-\epsilon,
\end{equation}
where the frequencies, $\omega_k$, and displacements, $\xi_k$, are obtained by discretising a spectral density according to
\begin{equation}
  J(\omega) = \frac{\pi}{2}\sum_{k=1}^f \frac{c_k^2}{m\omega_k}\delta(\omega-\omega_k)
\end{equation}
with $c_k=m\omega_k^2\xi_k$.

In the nonadiabatic limit the spin-boson problem is exactly soluble. The correlation function $c(t)=e^{-\phi(t)/\hbar}$ can be expressed in terms of the spectral density $J(\omega)$ as\cite{Weiss08}
\begin{equation}
  \begin{aligned}
  \phi(t)&=\phi\bigg(-\frac{i\beta\hbar}{2}\bigg)-\bigg(\frac{\beta\hbar}{2}-it\bigg)\epsilon \\ &+ \frac{4}{\pi}\int_0^{\infty}\frac{J(\omega)}{\omega^2}\bigg(\frac{\cosh\frac{1}{2}\beta\hbar\omega-\cos\omega t}{\sinh\frac{1}{2}\beta\hbar\omega}\bigg)\mathrm{d}\omega,
\end{aligned}
\end{equation}
in which the first term gives $e^{-\phi(-i\beta\hbar/2)/\hbar} = Q_{\rm r}(T)$.
Hence the rate can be calculated exactly for this problem by evaluating the time integral in Eq.~(7) numerically. This is typically done by shifting the contour of integration to pass through the saddle point at $t_{\rm sp}=i\lambda_{\rm sp}\hbar$, which makes the integrand less oscillatory.

Using the relation $\phi(i\lambda\hbar)/\hbar=\beta [F(\lambda)-({1/2}+{\lambda/\beta})\epsilon]$, we also have that, in its discretised form, $ F_{\epsilon}(\lambda)$ in Eq.~(28)  is given exactly by
\begin{equation}
  \begin{aligned}
   F_{\epsilon}(\lambda) &= -\bigg(\frac{1}{2}+\frac{\lambda}{\beta}\bigg)\epsilon \\ &+ \sum_{k=1}^f \frac{2c_k\xi_k}{\beta\hbar\omega_k}\bigg(\frac{\cosh\frac{1}{2}\beta\hbar\omega_k-\cosh\lambda\hbar\omega_k }{\sinh\frac{1}{2}\beta\hbar\omega_k}\bigg).
  \label{exact_free_energy}
\end{aligned}
\end{equation}
This suggests an obvious ansatz for $ \tilde{F}_{\epsilon}(\lambda)$ of the form
\begin{equation}
  \begin{aligned}
   \tilde{F}_{\epsilon}(\lambda) &= -\bigg(\frac{1}{2}+\frac{\lambda}{\beta}\bigg)\epsilon \\
  &+\sum_{i=1}^N \frac{a_i}{b_i^2}\left(\cosh {\textstyle{\frac{1}{2}}}b_i \beta -\cosh b_i\lambda \right),
  \label{spin_boson_ansatz}
  \end{aligned}
\end{equation}
in which $\{a_i\}$ and $\{b_i\}$ are real fitting parameters with $a_i>0$. Eq.~(35) is clearly capable of giving an accurate fit to the path integral data when $N=f$. However, we expect that it will also give a good fit that can be extrapolated some way outside $-\beta/2\leq\lambda\leq\beta/2$ for $N\ll f$. This is because the exact function contains information about the behaviour of $c(t)$ well away from the imaginary time axis, which is not important for the evaluation of Eq.~(27).

Given the success of the spin-boson model in describing real physical systems,\cite{Chandler89} we expect that this ansatz will also work well for other ``symmetric'' problems in which the reactant and product states are degenerate when $\epsilon=0$ and $F(\lambda)$ is an even function of $\lambda$. Systems which do not have this property will require an extension of the ansatz, and this is discussed in Sec.~IV.B.

\begin{figure}[t]
\centering
\resizebox{0.95\columnwidth}{!} {\includegraphics{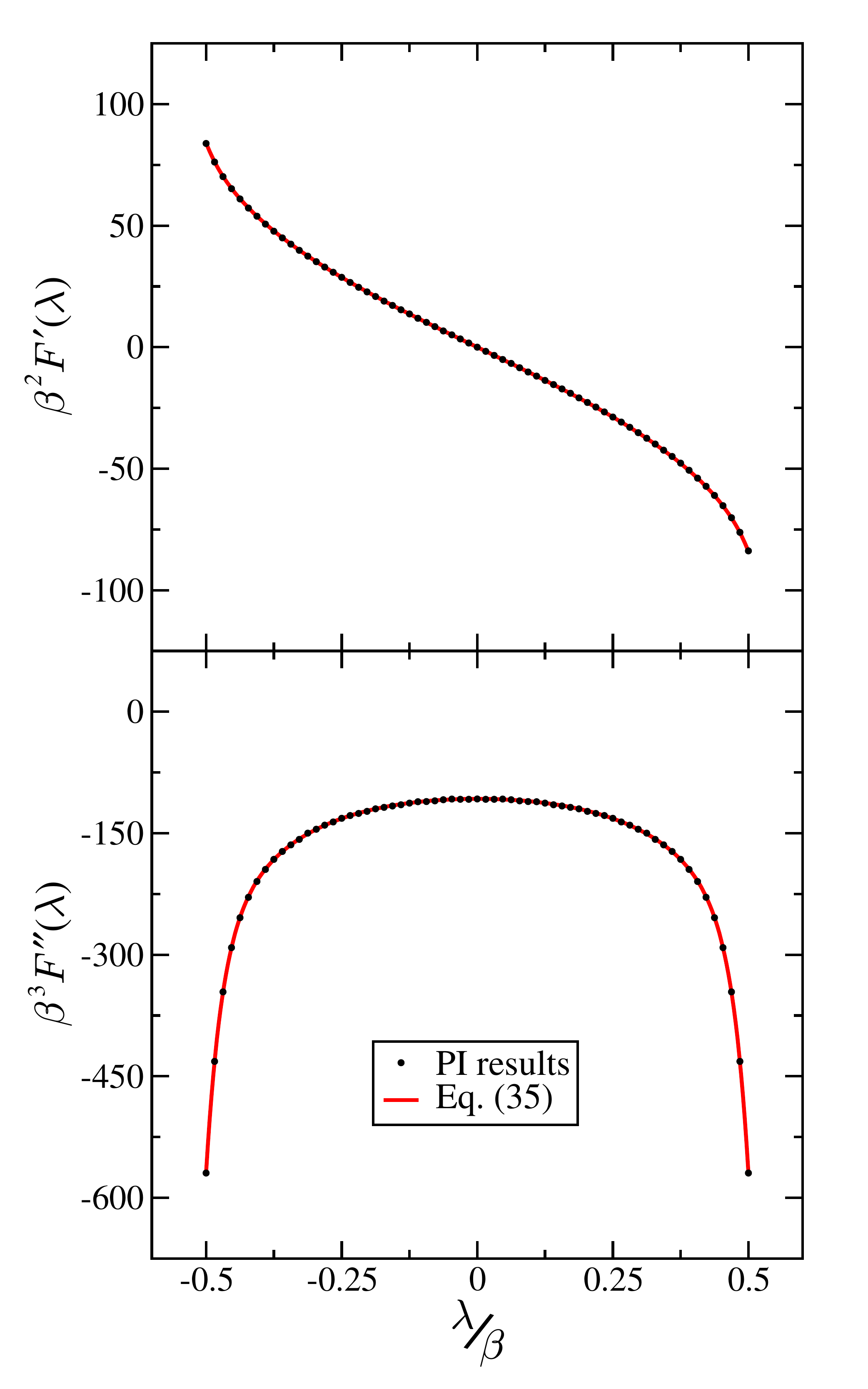}}
\caption{Comparison of the path integral data for the spin-boson model with $\epsilon=0$ and the fit obtained to it using Eq.~(35) with $N=3$. The standard errors in the path integral results are smaller than the size of the dots.} 
\label{SB_fits_figure}
\end{figure}
  
\begin{figure*}[t]
\centering
\resizebox{0.7\textwidth}{!} {\includegraphics{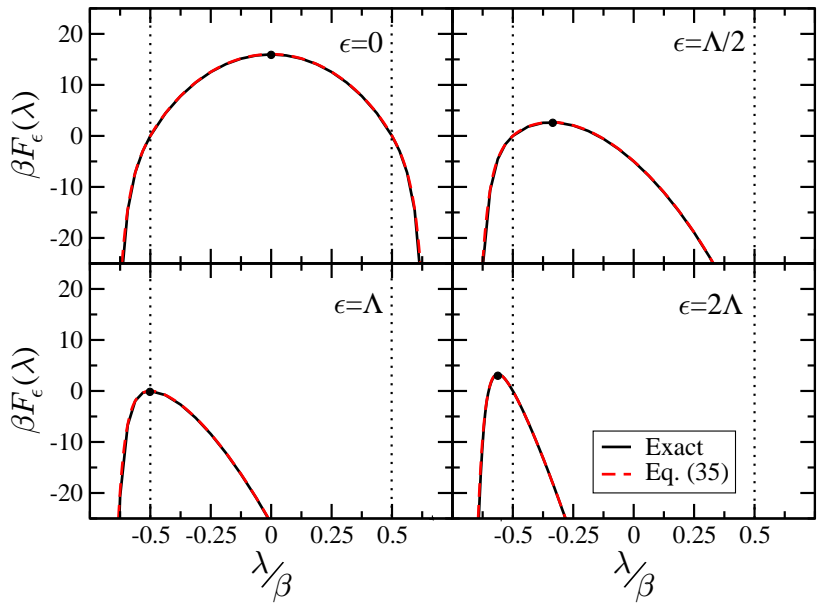}}
\caption{A comparison of the exact $F_{\epsilon}(\lambda)$ [calculated using Eq.~(34)] and the result of fitting the ansatz in Eq.~(35) to path integral data, for the spin-boson model. The solid black circle in each panel shows the saddle point, which moves from $\lambda_{\rm sp}=0$ for symmetric electron transfer ($\epsilon=0$) to $\lambda_{\rm sp}=-\beta/2$ for activationless electron transfer ($\epsilon=\Lambda$) to $\lambda_{\rm sp}<-\beta/2$ in the Marcus inverted regime ($\epsilon>\Lambda$).}
\label{SB_F_lambdas}
\end{figure*}

To test the method described in Sec.~III, we have considered a spin-boson model with Debye spectral density
 \begin{equation}
   J(\omega) = \frac{\Lambda}{2}\frac{\omega\omega_c}{\omega^2+\omega_c^2},
 \end{equation}
 with reorganisation energy $\Lambda=50\,\mathrm{kcal}/{\mathrm{mol}}$ and characteristic frequency $\omega_c=500\,\mathrm{cm}^{-1}$, at $300\,\mathrm{K}$. These parameters are similar to those used in several earlier studies and they were chosen here because they give a significant quantum mechanical effect on the rate constant: they lead to a rate constant for symmetric electron transfer ($\epsilon=0$) around two orders of magnitude larger than the Marcus theory prediction, which is at the upper limit of earlier estimates of the quantum enhancement for the ferrous-ferric system in aqueous solution.\cite{Bader90,Blumberger08} 

The spectral density was discretised in the standard way, with\cite{Richardson15b}
 \begin{equation}
   \omega_k = \omega_c \tan \frac{(k-\frac{1}{2})\pi}{2f},
 \end{equation}
 and
 \begin{equation}
   c_k=\sqrt{\frac{m \Lambda}{2f}}\omega_k,
 \end{equation}
 and 
 \begin{equation}
 \xi_k=\sqrt{\Lambda\over 2mf}{1\over\omega_k},
 \end{equation}
 for $k=1,\ldots,f$. (Note that since neither $\omega_k$ nor $c_k\xi_k$ in Eq.~(34) depends on $m$ this is a redundant parameter that does not need to be specified to define the problem.) It was found that $f=12$ degrees of freedom were sufficient to converge the exact rate to graphical accuracy for all $\epsilon$ considered.

The parameters $a_i$ and $b_i$ were determined by simultaneously fitting $\{F'(\lambda_j)\}$ and $\{F''(\lambda_j)\}$ to the ansatz in Eq.~(35) with $\epsilon=0$. This was done by minimising the objective function
\begin{equation}
  \begin{aligned}
  L(\mathbf{a},\mathbf{b}) &= \frac{\sum_j\big[ \tilde{F}_{0}'(\lambda_j)-F'(\lambda_j)\big]^2}{2\sum_j\big[\bar{F}'-F'(\lambda_j)\big]^2} \\&+ \frac{\sum_j\big[ \tilde{F}_{0}''(\lambda_j)-F''(\lambda_j)\big]^2}{2\sum_j\big[\bar{F}''-F''(\lambda_j)\big]^2},
\end{aligned}
\end{equation}
where $\bar{F}'=\frac{1}{N_j}\sum_j F'(\lambda_j)$ and $\bar{F}''=\frac{1}{N_j}\sum_j F''(\lambda_j)$ with $N_j$ being the number of $j$ values at which the derivatives were calculated.

As this is a non-linear optimisation problem there may be many local minima. We therefore used a global optimisation algorithm\cite{nlopt,nloptISRES} on the domain defined by $0\leq a_i\leq |F''(0)|$ and $0< b_i\leq\hbar\omega_{\mathrm{max}}$, with $\omega_{\mathrm{max}}=22000\,\mathrm{cm}^{-1}$. The path integral results were calculated using path integral molecular dynamics with the PILE thermostat\cite{Ceriotti10} at 65 equally spaced points in the interval $-\beta/2\leq\lambda_j\leq\beta/2$, with $n=256$ path integral beads. We found that $N=3$, corresponding to 6 free parameters, was sufficient to fit the results to within their statistical error bars. Fig.~3 shows the first and second derivatives of $F(\lambda)$ along with the corresponding fit obtained for $N=3$, for which  $\beta^3\mathbf{a}=( 1.4205\times10^{-5},  1.0787\times10^{1},    9.6756\times10^{1})$ and $\beta\mathbf{b}=(3.4972\times10^{1}, 7.0368, 9.5562\times10^{-1}  )$.

\begin{figure}[t]
\centering
\resizebox{0.95\columnwidth}{!} {\includegraphics{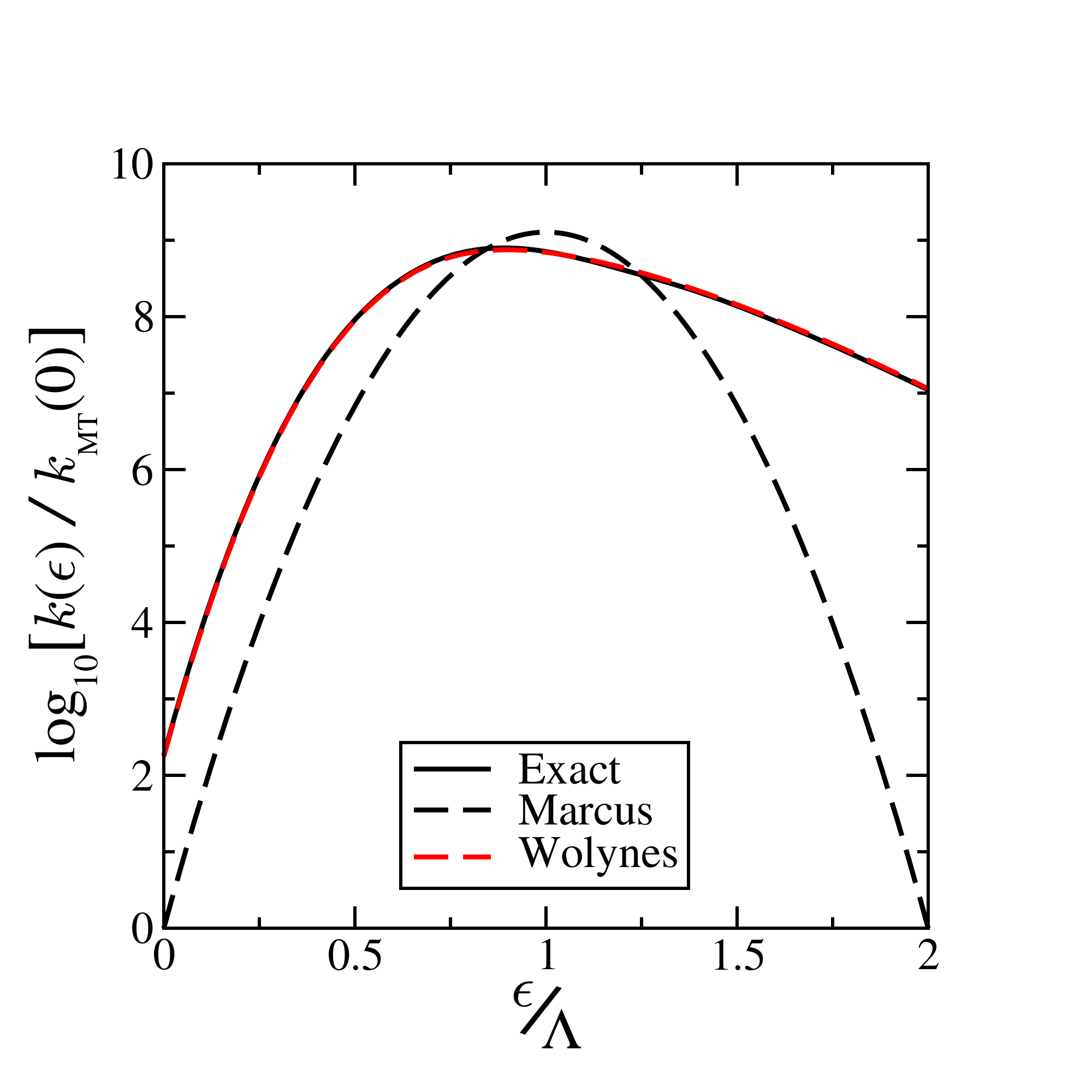}}
\caption{Rate constants for the spin-boson model, relative to the Marcus theory result for $\epsilon=0$. The exact results were calculated using Eq.~(7) and the Wolynes results were calculated using Eq.~(27).}
\label{SB_rates}
\end{figure}

As $F_{\epsilon}(\lambda)$ is known analytically for this problem we can compare the exact curve to the result of fitting to the path integral data. Fig.~4 shows $ \tilde{F}_{\epsilon}(\lambda)$ (with the parameters given above) and the exact $F_{\epsilon}(\lambda)$ for a variety of driving forces, ranging from the normal regime to activationless electron transfer and on into the Marcus inverted regime. The agreement is clearly excellent even for $\epsilon=2\Lambda$, deep inside the inverted regime. This justifies our assertion that the behaviour on the imaginary axis can be captured by including only a small number of terms $N\ll f$ in Eq.~(35). (One could argue that this implies that an alternative discretisation of $J(\omega)$ would have led to a faster convergence with respect to $f$. However, without actually doing the path integral calculations and fitting the results to Eq.~(35) there would be no \emph{a priori} way of knowing what the appropriate effective frequencies and couplings were.)

Having found the parameters in $\tilde{F}_{\epsilon}(\lambda)$, we are now in a position to evaluate Eq.~(27) for arbitrary $\epsilon$, as described in Sec.~III. Fig.~5 compares the exact rate constants to those calculated using Eq.~(27) with $F_{\epsilon}(\lambda)$ replaced by $\tilde{F}_{\epsilon}(\lambda)$. The exact rate constants were obtained by substituting Eq.~(33) into Eq.~(7) and doing the time integral numerically. For comparison the rate constants predicted by Marcus theory, which is exact in the classical ($\beta\to0$) limit for this problem, are given by\cite{Marcus85}
\begin{equation}
  k_{\rm MT}(T) = \frac{\Delta^2}{\hbar}\sqrt{\frac{\pi\beta}{\Lambda}}e^{-\beta(\Lambda-\epsilon)^2/4\Lambda}.
\end{equation}
Since all the rate constants in Fig.~5 are given relative to the classical rate at $\epsilon=0$ we do not need to specify the value of $\Delta$. However,  for the results to be accurate $\Delta$ must be small enough for the golden rule to apply. It is apparent that Wolynes theory gives excellent agreement with the exact rate for all physically relevant values of $\epsilon$. Note that this is a situation in which nuclear quantum effects are vitally important: the quantum mechanical rate is 7 orders of magnitude larger than the classical rate at $\epsilon=2\Lambda$. The success of the ansatz in Eq.~(35) for this problem is striking and suggests that the same approach is likely to work for a wide variety of ``symmetric'' condensed phase electron transfer reactions. The extension of the approach to more general (asymmetric and anharmonic) problems is discussed next.

\subsection{An electronic predissociation model}

The second system we shall consider is a one-dimensional model of electronic predissociation that was introduced by Richardson and Thoss to illustrate the oscillatory nature of nonadiabatic reactive flux correlation functions.\cite{Richardson14} Analogous to a condensed phase problem, this system has a continuum of product states, and it provides a useful test case because the exact nonadiabatic rate constant can again be calculated for comparison (in this case using quantum scattering theory). 

The diabatic potential energy curves of the model are
\begin{equation}
  V_0(q) = \frac{1}{2}m\omega^2q^2,
\end{equation}
and
\begin{equation}
  V_1(q) = De^{-2\alpha(q-\xi)}-\epsilon,
\end{equation}
in which we shall use the same parameters as Richardson and Thoss:\cite{Richardson14} $m=1$, $\omega=1$, $D=2$, $\alpha=0.2$, $\xi=5$, $\beta=3$, and $\hbar=1$. Note that for this problem the reorganisation energy as defined (at $\epsilon=0$) in Eq.~(29) is temperature-dependent, because although $V_0(q)$ is still harmonic $V_1(q)-V_0(q)$ is no longer a linear function of $q$. When presenting our results as a function of $\epsilon/\Lambda$ we shall therefore simply define $\Lambda$ as $\Lambda=De^{2\alpha\xi}$, as illustrated in Fig.~6. 

\begin{figure}[t]
\centering
\resizebox{0.95\columnwidth}{!} {\includegraphics{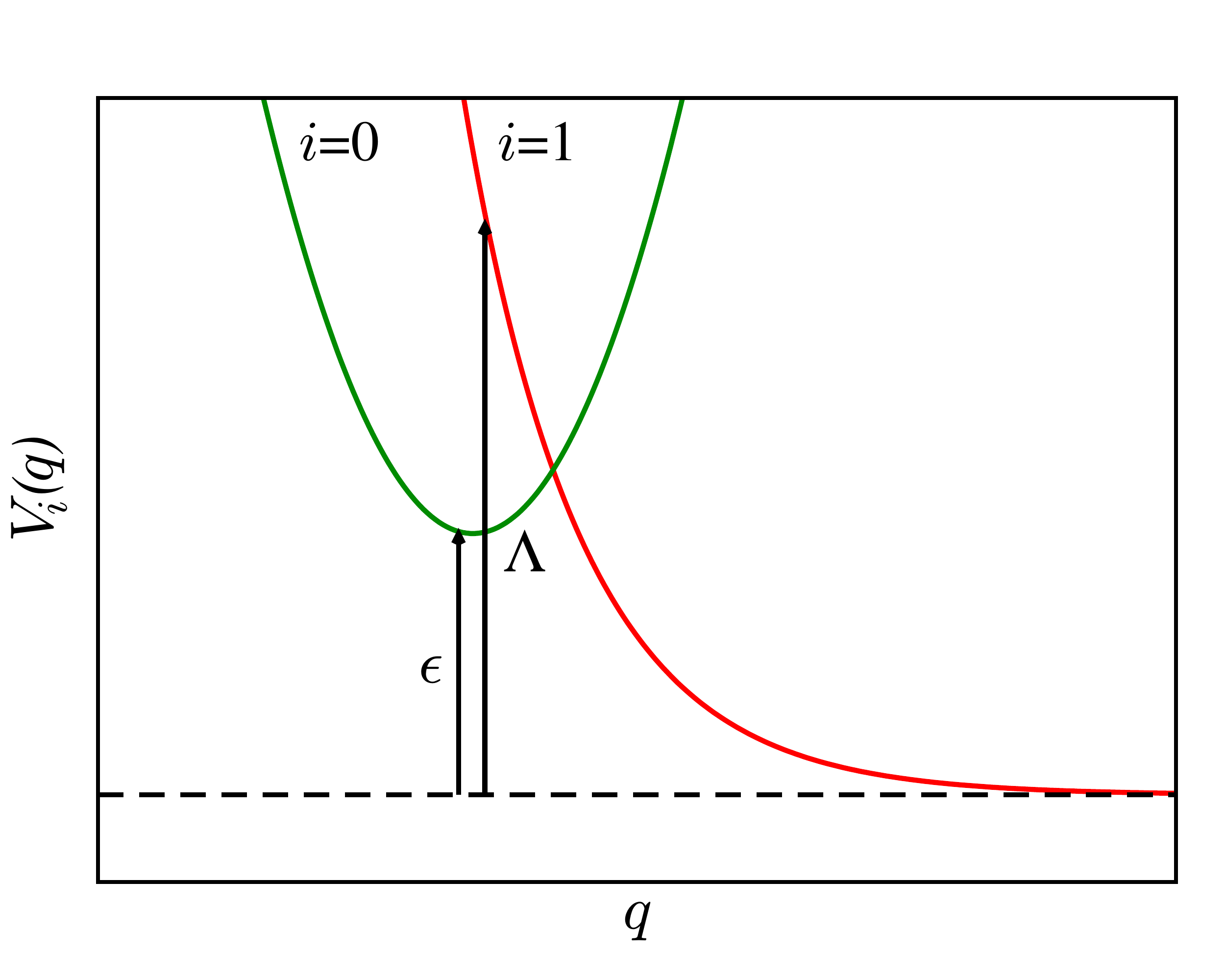}}
\caption{The diabatic potential energy curves for the one-dimensional electronic predissociation model, along with the driving force $\epsilon$ and reorganisation energy $\Lambda$. }
\label{PreDisDiagram}
\end{figure}

This model presents two obvious challenges which will require modifications of the ansatz for $ \tilde{F}_{\epsilon}(\lambda)$ in Eq.~(35). Firstly the system is not symmetric and secondly, since $V_1(q)$ is unbounded, the product partition function is infinite and the imaginary time correlation function $c(i\lambda\hbar)$ diverges at $\lambda=\beta/2$. Assuming a power law divergence,
\begin{equation}
  c(i\lambda\hbar) \sim A(\beta/2-\lambda)^{-\gamma} \text{ as $\lambda\to\beta/2$} 
\end{equation}
for some positive constant $\gamma$, it follows from the definition of $F(\lambda)$ that
\begin{equation}
  F(\lambda)\sim\frac{\gamma}{\beta}\ln(\beta/2-\lambda) \text{ as $\lambda\to\beta/2$}.
\end{equation}

On the basis of these observations, we arrive at the following generalisation of Eq.~(35):
\begin{equation}
  \begin{aligned}
   \tilde{F}_{\epsilon}(\lambda) &= d \ln\left({1\over 2}-{\lambda\over\beta}\right)-\left({1\over 2}+{\lambda\over\beta}\right)\epsilon \\
  &+ \sum_{i=1}^N \frac{a_i}{b_i^2}\left(\cosh b_i\left[\textstyle{1\over 2}\beta+c_i\right] - \cosh b_i\bigl[\lambda-c_i\bigr]\right). 
  \end{aligned}
\end{equation}
Here again $a_i>0$, and in addition to the logarithmic term we have accounted for asymmetry in $F(\lambda)$ by adding the additional parameters $c_i$ which shift the origins of the hyperbolic cosines.

The parameters were again determined by minimising the objective function
\begin{equation}
  \begin{aligned}
  L(\mathbf{a},\mathbf{b},\mathbf{c},d) &= \frac{\sum_j\big[ \tilde{F}_{0}'(\lambda_j)-F'(\lambda_j)\big]^2}{2\sum_j\big[\bar{F}'-F'(\lambda_j)\big]^2} \\&+ \frac{\sum_j\big[ \tilde{F}_{0}''(\lambda_j)-F''(\lambda_j)\big]^2}{2\sum_j\big[\bar{F}''-F''(\lambda_j)\big]^2},
\end{aligned}
\end{equation}
with the same non-linear optimisation algorithm.\cite{nlopt,nloptISRES} The search was performed on the domain defined by $0\leq a_i\leq |F''(0)|$, $0< b_i\leq\hbar\omega_{\mathrm{max}}$ (with $\omega_{\mathrm{max}}=10$), $-\beta<c_i<\beta$, and $0\leq d\leq 15/\beta$.
The path integral results were calculated in the same way as for the spin-boson model at 32 equally spaced points in the interval $-\beta/2\leq\lambda_j\leq15\beta/32$, with $n=256$ path integral beads. $\lambda_n=\beta/2$ was not included because $F(\beta/2)$ is undefined.

We found that $N=3$ in Eq.~(46) was again sufficient to fit the path integral results to within their statistical error bars.
Fig.~7 shows the first and second derivatives of $F(\lambda)$ along with the corresponding fit, for which  $\beta^3\mathbf{a}=( 1.8281\times10^{1},\,  3.4729\times10^{-2},\,    2.8976\times10^{1})$, $\beta\mathbf{b}=(8.4376\times 10^{-3},\,1.9063\times10^{1},\, 4.6062)$, $\mathbf{c}/\beta=(7.4223\times10^{-1},\,-6.1285\times10^{-2},\,-8.9571\times10^{-2})$ and $\beta d=1.3762$.

\begin{figure}[b]
\centering
\resizebox{0.95\columnwidth}{!} {\includegraphics{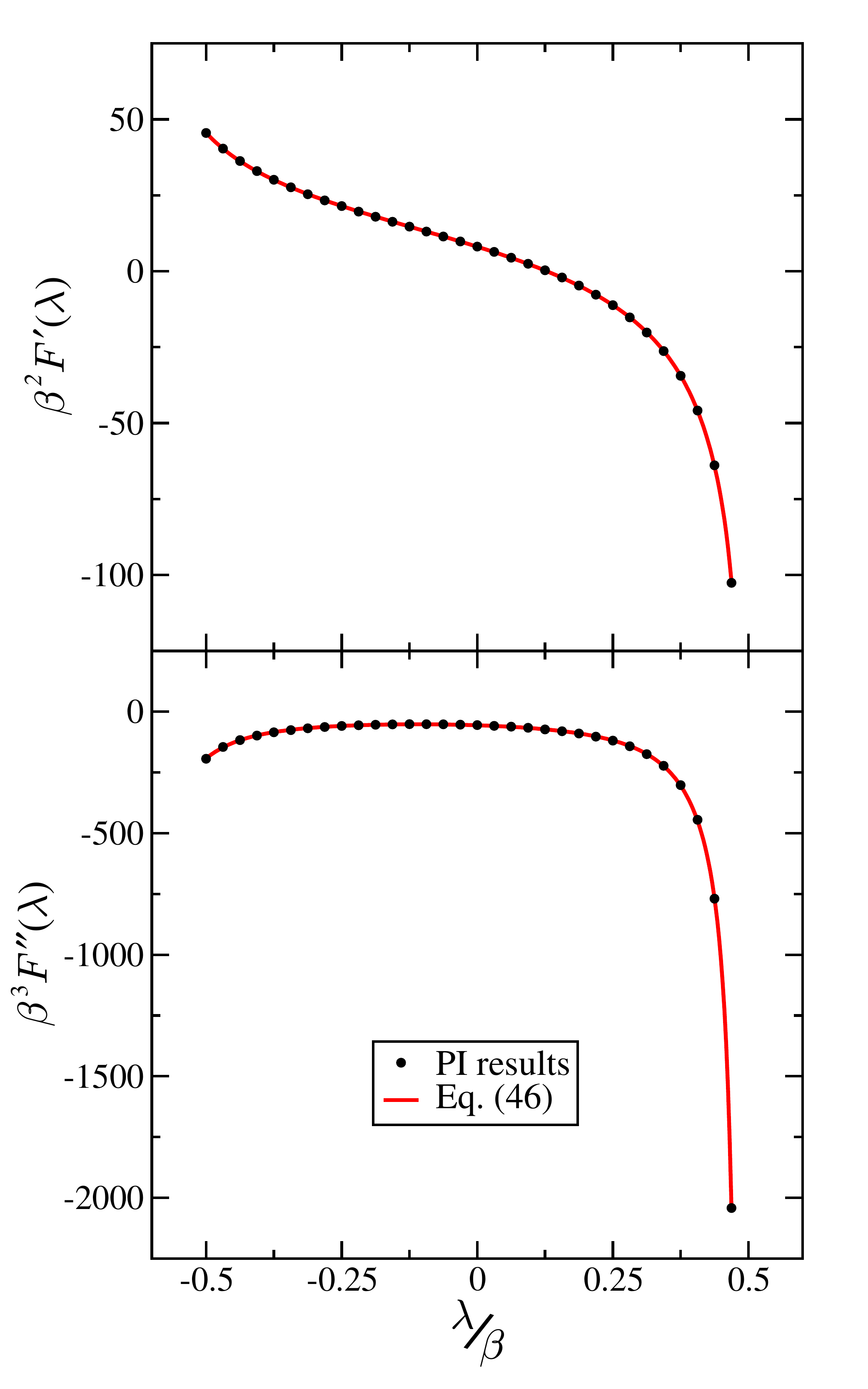}}
\caption{Comparison of the path integral data for the electronic predissociation model with $\epsilon=0$ and the fit obtained to it using Eq.~(46) with $N=3$. The standard errors in the path integral results are smaller than the size of the dots.}
\label{PreDisFits}
\end{figure}

\begin{figure}[h]
\centering
\resizebox{0.95\columnwidth}{!} {\includegraphics{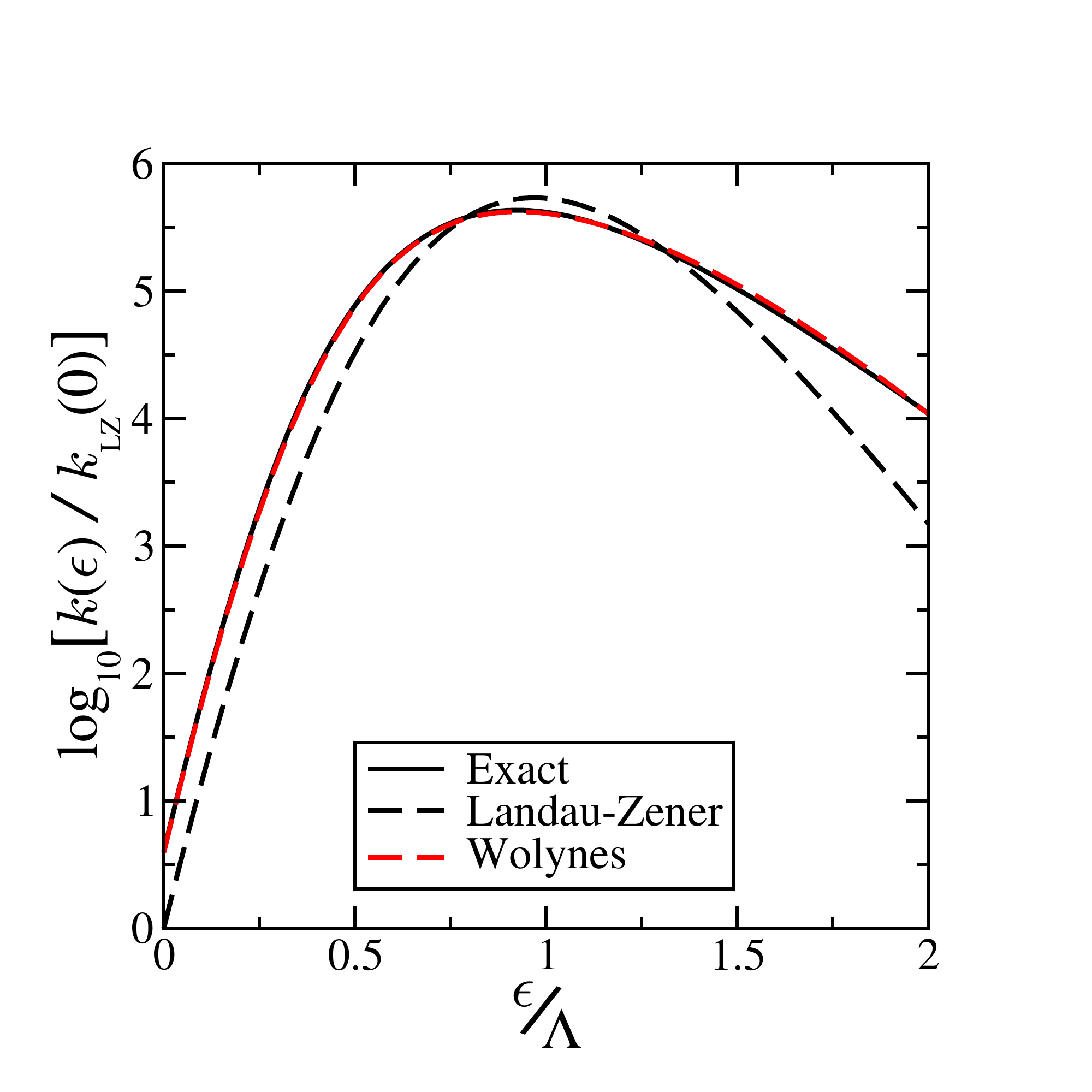}}
\caption{Rate constants for the electronic predissociation model, relative to the Landau-Zener result for $\epsilon=0$. The exact results were calculated using Eq.~(48) and the Wolynes results were calculated using Eq.~(27).}
\label{PreDisRates}
\end{figure}

We have also computed the exact golden rule rate for this problem for comparison. This was done by using a Lobatto shape function discrete variable representation\cite{Manolopoulos88} to calculate
\begin{equation}
  k(T)Q_{\rm r}(T) = - \frac{2\Delta^2}{\hbar}\sum_\nu e^{-\beta E_\nu} \Im\bra{\nu}\hat{G}^+_1(E_\nu)\ket{\nu},
  \label{scattering_rate}
\end{equation}
and
\begin{equation}
Q_{\rm r}(T) = \sum_{\nu} e^{-\beta E_{\nu}},
\end{equation}
where 
\begin{equation}
\hat{H}_0\ket{\nu}=E_\nu\ket{\nu},
\end{equation}
and
\begin{equation}
\hat{G}^+_1(E_\nu) = \lim_{\eta\to 0_+}\big(E_\nu+i\eta-\hat{H}_1\big)^{-1}.
\end{equation}

For this problem the appropriate classical limit of the rate constant is given by the Landau-Zener formula\cite{Nitzan06}
\begin{equation}
  k_{\rm LZ}(T) = \frac{\Delta^2}{\hbar}\sqrt{2\pi\beta m\omega^2}\frac{e^{-\beta V_0(q^\ddagger)}}{|V'_0(q^\ddagger)-V'_1(q^\ddagger)|},
\end{equation}
in which $q^\ddagger$ is defined by the equation $V_0(q^\ddagger)=V_1(q^\ddagger)$ and we have included a factor of 2 in the prefactor to allow for transitions with both positive and negative momenta. 

Fig.~8 compares the exact, Wolynes and Landau-Zener rate constants for this problem relative to $k_{\rm LZ}(\epsilon=0)$. Once again Wolynes theory gives excellent agreement with the exact rate well into the Marcus inverted regime. The quantum mechanical enhancement of the rate is not as pronounced as it was for the spin-boson model considered in Sec.~IV.A, but it is still almost an order of magnitude at $\epsilon=2\Lambda$. And this problem is more challenging in many other respects, including its asymmetry, its anharmonicity, and the divergent behaviour of its free energy as $\lambda\to\beta/2$. We therefore regard the results in Fig.~8 to be just as encouraging for future applications of the theory as those in Fig.~5.
\\

\section{Conclusions and Future Work}

In this paper, we have established that an appropriate implementation of Wolynes theory\cite{Wolynes87} works equally well in both the normal and inverted electron transfer regimes. The only difference between the two regimes is that, whereas the treatment of the normal regime involves the interpolation of imaginary time path integral data, the treatment of the inverted regime involves its extrapolation. While extrapolation can often be problematic, this is not the case in the present context, because one does not have to extrapolate very far along the imaginary time axis to reach the saddle point that determines the Wolynes rate. This is certainly true for both of the model problems we have considered -- a typical spin-boson problem and a one-dimensional model of electronic predissociation -- and we would expect it to be true more generally. 

Given this, and the fact that the imaginary time path integral data is straightforward to calculate for arbitrarily complex systems (there is no sign problem and the scaling of the calculation is linear in the number of degrees of freedom), we feel that Wolynes theory has not received the attention it deserves. When compared with many of the more recent alternative theories, including in particular those based on the semiclassical instanton approximation,\cite{Richardson15a,Richardson15b} and those based on various electronically nonadiabatic generalisations of ring polymer molecular dynamics,\cite{Shushkov12,Richardson13,Ananth13,Duke16,Menzeleev14,Kretchmer16} Wolynes theory seems to us to provide a very simple, practical, and reliable way to include quantum mechanical effects in the nuclear motion in the calculation of electronically nonadiabatic reaction rates.

We therefore intend to explore the theory further in future work, and to apply it to a variety of electron transfer reactions in chemical and biochemical systems. There are a number of interesting aspects to this, such as just how important nuclear quantum effects really are in these systems, how reliable it is to map a simulation of an electron transfer reaction with anharmonic reactant and product states onto a harmonic spin-boson model in order to calculate the electron transfer rate,\cite{Ando01,Blumberger06,Blumberger08} how well the nonadiabatic to adiabatic transition can be captured by going beyond the golden rule limit and including contributions to $c(t)$ of higher order in $\Delta$,\cite{Wolynes87,Mukamel88} and whether one can extract any mechanistic information from the path integral simulations at each $\lambda_j$ about what is happening to the nuclei as the electron transfers from the reactants ($\lambda_j=-\beta/2$), through the electron transfer transition state ($\lambda_j\simeq \lambda_{\rm sp}$), to the products ($\lambda_j=+\beta/2$). 

\begin{acknowledgments}
We are grateful to Damien Laage for helpful discussions during the early stages of this work and to Jeremy Richardson for suggesting the electronic predissociation model considered in Sec.~IV.B. J. E. Lawrence is supported by The Queen's College Cyril and Phillis Long Scholarship in conjunction with the Clarendon Fund of the University of Oxford and by the EPRSC Centre for Doctoral Training in Theory and Modelling in the Chemical Sciences, EPSRC grant no. EP/L015722/1.
\end{acknowledgments}


\begin{thebibliography}{100}
\bibitem{Footnote1} Alternatively, one can analytically continue the path integral data back to the real time axis by setting $\lambda=-it/\hbar$ in our $\tilde{F}_{\epsilon}(\lambda)$, and integrate the resulting correlation function $c(t)=Q_r(T)e^{-\beta\tilde{F}_{\epsilon}(-it/\hbar)}$ numerically over $t$, thereby avoiding any phase integral approximation. We have done this for both of the model problems we have studied, and found that the \lq\lq Wolynes" results in Figs.~5 and~8 were unchanged to graphical accuracy: the saddle point approximation is perfectly adequate for both of these problems. (We have based the presentation in the text on this approximation rather than the numerical integration of the analytically-continued correlation function because it is simpler to implement, and because it makes it clear that the path integral data does not have to be continued very far along the imaginary time axis to calculate the rate in the inverted regime.)
\bibitem{Footnote2} It is however conceivable that the instanton approximation could be extended into the Marcus inverted regime using a similar procedure to that outlined in the present paper.
\end{thebibliography}
\end{document}